\magnification=\magstep1
\tolerance=500
\vskip 2 true cm
\rightline{TAUP 2837/06}
\rightline{27 March, 2007}
\bigskip
\centerline{\bf On The Geometry of Hamiltonian Chaos}
\bigskip
\centerline{Lawrence Horwitz$^{1,2,3}$,Jacob Levitan$^{1,4}$,
Meir Lewkowicz$^1$,  Marcelo Schiffer$^1$ and Yossi Ben Zion$^{1,3}$}
\centerline{$^1$Department of Physics, College of Judea and Samaria,
  Ariel 44837, Israel}
\centerline{$^2$School of Physics, Tel Aviv University, Ramat Aviv 69978,
  Israel}
\centerline{$^3$Department of Physics, Bar Ilan University, Ramat
Gan 52900, Israel}
\centerline{$^4$ Department of Physics, Technical University of Denmark,
Lyngby 2800, Denmark}
\bigskip
\bigskip
\noindent {\it Abstract:\/} The characterization
of chaotic Hamiltonian
systems in terms of the curvature associated with a Riemannian metric
tensor in the structure of the Hamiltonian can be extended to a wide
class of potential models of standard form through definition of a
conformal metric. The geodesic equations reproduce the
Hamilton equations of the original potential model when a transition
is made to an associated manifold for which the geodesics coincide with the 
orbits of the Hamiltonian potential model. We
therefore find a direct geometrical
description of the time development of a Hamiltonian potential model.
The second covariant derivative of the geodesic deviation in
 this associated manifold generates a dynamical curvature, resulting in
(energy dependent) criteria for unstable behavior different
from the usual Lyapunov criteria. We discuss some examples of unstable
Hamiltonian systems in two dimensions giving, as a particular illustation,
detailed results for a potential obtained from a fifth order expansion
of a Toda lattice Hamiltonian.
\smallskip
\leftline{ PACS: 45.20.Jj, 47.10.Df, 05.45.-a, 05.45.Gg}
\bigskip
\bigskip
\par A Hamiltonian system of the
form (we use the summation convention)
$$ H = {1 \over 2M} g_{ij}p^i p^j, \eqno(1)$$
where $g_{ij}$ is a function of the coordinates alone, has unstable orbits if
the curvature associated with the metric $g_{ij}$ is
negative. One can easily see that the orbits described by the Hamilton
equations for $(1)$ coincide with the geodesics on a Riemannian space 
associated with the metric $g_{ij}$, $^{1,2}$ i.e.,
it follows directly from the Hamilton equations associated with
$(1)$ that (using $(12)$ and the time derivative of $(10)$) 
$$ {\ddot x}_\ell = -\Gamma_\ell^{mn} {\dot x}_m {\dot x}_n,
\eqno(2) $$
where the connection form $\Gamma_\ell^{mn}$ is given by
$$ \Gamma_\ell^{mn} = {1\over 2} g_{\ell k} \bigl\{ {\partial g^{km} \over
\partial x_n}+  {\partial g^{k n} \over
\partial x_m}- {\partial g^{n m} \over
\partial x_k} \bigr\}, \eqno(3)$$
and $g^{ij}$ is the inverse of $g_{ij}$.
\par The second covariant derivative of  the geodesic deviation depends on
the curvature$^{2,3}$
$$ R_i^{jk\ell} = {\partial \Gamma_i^{jk} \over \partial
x_\ell} - {\partial \Gamma_i^{j\ell} \over \partial x_k}
+ \Gamma_m^{jk} \Gamma_i^{\ell m} - \Gamma_m^{j\ell} \Gamma_i^{km};
\eqno(4)$$
 i.e., for $\xi_i =
x'_i -x_i$ on closely neighboring trajectories at $t$,
$$ {D^2 \xi_i \over Dt^2} =  R_i^{j\ell k} {\dot x}_j {\dot x}_k
\xi_\ell, \eqno(5) $$
where $D/Dt$ is the covariant derivative along the line $x_j(t)$. The
sign of the scalar contraction of $(4)$ then gives information on the
stability of the orbits.$^3$
 \par In this Letter, we
point out that this formulation of dynamic stability has application
to a much wider range of Hamiltonian models; in fact, every potential
model Hamiltonian
of the form
$$ H= {{p^i}^2 \over 2M} + V(x), \eqno(6)$$
where $V$ is a function of space variables alone, can be put into the
form $(1)$, where the metric tensor is of conformal form.$^4$
 We obtain in this way a direct geometrical
description of the time development for a Hamiltonian potential
model.
\par Casetti, Pettini and collaborators$^5$, for example, have studied
the application of both the Jacobi and Eisenhardt metrics in their
analyses of the
geometry of Hamiltonian chaos.  The Jacobi metric$^1$ (of the form
$(E-V)\delta_{ij} $) leads to geodesic equations parametrized by the
invariant distance associated with this metric on the manifold, in
this case, the kinetic energy, thus
corresponding to the Hamilton action.  Transformation to
parametrization by the time $t$ leads to the second order Newton
law$^5$ in the form $(14)$ below,  for which the geometrical structure
is no longer evident.
\par  The Eisenhardt metric, leading to geodesic motion in $t$,
involves the addition of an extra dimension. As noted by Caini {\it et
 al\/}$^5$, this metric leads to the tangent dynamics commonly used to
 measure Lyapunov exponents in standard Hamiltonian systems. The
 method that we use, associated with a curvature that is explicitly
 energy dependent, appears to be a more sensitive diagnostic than the
 computation of exponents of a locally linearized system. 
\par  The formulation of Hamiltonian dynamics of the type of Eq.$(6)$
in the form $(1)$ is carried out by requiring that $(6)$ be equivalent
to $(1)$. For a metric of conformal form
$$g_{ij} = \varphi \delta_{ij}, \eqno(7)$$
on the hypersurface defined by $H=E=constant$, the requirement of
equivalence implies that
$$ \varphi = {E \over E-V(x)}.\eqno(8)$$
\par Substituting this result in the geodesic equations $(2)$, one
obtains an equation that does not coincide in form with the Hamilton
equations obtained from $(6)$.
\par To see that the Hamilton equations obtained from $(1)$ can,
however,  be put
 into correpondence with those obtained from the Hamiltonian of the potential
model $(6)$, we first note, from the Hamilton equations for $(1)$, that
$$ {\dot x}_i = {\partial H \over \partial p^i}= {1 \over M}g_{ij}
 p^j. \eqno(9)$$
We then use the geometrical property that ${\dot x}_i$ is a first rank
tensor (as is $p^i$), under local
 diffeomorphisms that preserve the constraint that $H$ be constant, to
define the {\it velocity field}
$$ {\dot x}^j \equiv  g^{ji}{\dot x}_i = {1 \over M}p^j, \eqno(10)$$
coinciding formally with one of the Hamilton equations implied by
$(6)$. From this definition, we recognize that we are dealing with two
manifolds, each characterized, as we shall see, by a different
connection form, but related by
$$ dx^j = g^{ji}dx_i \eqno(11)$$
on a common tangent space at each point (for which $g^{ij}$ is nonsingular). 
\par To complete our correspondence with the dynamics induced by
$(6)$, consider the Hamilton equation for ${\dot p}^i$,
$$ {\dot p}^\ell = -{\partial H \over \partial  x_\ell}
= -{1 \over 2M}{\partial g_{ij} \over
\partial x_\ell}  p^i p^j .\eqno(12)$$
With the form $(7)$ for $g^{ij}$, we obtain in the particular
coordinate system in which $(6)$ is defined,
$$  {\dot p}^\ell = - {E \over E-V} {\partial V \over \partial x_\ell}. 
\eqno(13)$$
Considering $(11)$ as a change of variables,
 $(13)$  becomes
$$ {\dot p}^\ell = - {\partial V \over \partial x^\ell}, \eqno(14)$$
the second Hamilton equation in the usual form, where $V$ is
considered a function of the $\{ x^\ell\}$, now considered as
independent variables.
\par As a coordinate space, the $\{x^\ell\}$, which we shall call the
{\it Hamilton manifold}, is not uniquely defined
in terms of the original manifold $\{x_\ell\}$, which we shall call
the {\it Gutzwiller manifold}, since  $(11)$ is not an exact
differential. As we have remarked, we shall be working with two
manifolds (characterized by the connection forms $(3)$ and $(21)$).
 It is the local relation $(11)$ which induces,
from the geometry of the Gutzwiller manifold, a corresponding geometry
on the Hamilton manifold. We shall discuss applications and interpretation
of the physics of the Gutzwiller manifold elsewhere, but turn now to a further
examination of the consequences of the relations $(10)$ and $(111)$.
\par The geodesic equation $(2)$ can be transformed
directly from an equation for ${\ddot x}_j$ to an
 equation for ${\ddot x}^j$, the motion defined in the Hamilton
manifold. From $(10)$ it follows that
$$\eqalign{{\ddot x_\ell} &= g_{\ell j}{\ddot x}^j + {\partial g_{\ell j} \over
  \partial x_n} {\dot x}_n{\dot x}^j\cr &= - {1\over 2} g_{\ell k}
\bigl\{ {\partial g^{km} \over
\partial x_n}+  {\partial g^{k n} \over
\partial x_m}- {\partial g^{n m} \over
\partial x_k} \bigr\}{\dot x}_m{\dot x}_n .\cr} \eqno(15)$$
Now, using the identity
$$  {\partial g_{\ell j} \over
  \partial x_n} = -g_{\ell k} {\partial g^{km} \over \partial
x_n}g_{mj}, \eqno(16)$$
it follows that, with the symmetry of ${\dot x}_n{\dot x}_m$,
$$ {\partial g_{\ell j} \over
  \partial x_n} {\dot x}_n{\dot x}^j= -{1 \over 2}g_{\ell k} \bigl({\partial
g^{km} \over \partial x_n} +  {\partial
g^{kn} \over \partial x_m}\bigr){\dot x}_n {\dot x}_m. \eqno(17)$$
Thus, the term on the left side of $(15)$
containing the derivative of $g_{\ell j}$ cancels the
first two terms of the connection form; multiplying the result by the
inverse of $g_{\ell j}$, and applying the identity $(16)$ to lower the
indices of $g^{nm}$ in the remaining term on the right side of $(15)$,
one obtains 
$$  {\ddot x}^\ell = -M^\ell_{mn}{\dot x}^m {\dot x}^n, \eqno(18)$$
where
$$ M^\ell_{mn}\equiv {1 \over 2}g^{\ell k}{\partial g_{n m} \over
\partial x^k}. \eqno(19)$$
Eq. $(18)$ has the form of a geodesic equation, with a
truncated connection form.  In fact, it can be shown (a full proof
will be given elsewhere) that the
form $(19)$ is indeed a connection form, transforming as
$$ M'^\ell_{mn} = {\partial x'^\ell \over \partial x^r}
{\partial x^p \over \partial x'^m}{\partial x^q \over \partial x'^n}
M^r_{pq} + {\partial x'^\ell \over \partial x^r}{\partial^2 x^r \over
\partial x'^m \partial x'^n},$$
consistent with the covariance of $(18)$ under local diffeomorphisms
of the Hamilton manifold.
  \par  Substituting $(7)$ and $(8)$ into $(18)$ and $(19)$, the Kronecker
deltas identify the indices of ${\dot x}^m$ and ${\dot x}^n$; the
resulting square of the velocity cancels a factor of $(E-V)^{-1}$,
leaving the Hamilton-Newton law $(14)$.  Eq. $(18)$ is therefore a
covariant form of the Hamilton-Newton law, exhibiting what can be
considered an underlying geometry of standard Hamiltonian motion.
\par The geometrical structure of the Hamilton manifold
can be understood as follows. Let us write the covariant
derivative for a  (rank one) covariant tensor on the Gutzwiller manifold
 (defined as transforming in the
same way as $\partial/\partial x_m$), using the full connection form $(3)$,
$${A^m}^{;q} = {\partial A^m \over \partial x_q} - \Gamma_k^{mq} A^k .
 \eqno(20)$$
Lowering the index $q$ with $g_{\ell q}$, we obtain the covariant
 derivative in the Hamilton manifold, with connection form (with the
 help of $(16)$)
$$ \Gamma_{H\ell k}^m\equiv  g_{\ell q}\Gamma_k^{mq} = { 1 \over 2}g^{mq} 
\bigl\{ {\partial g_{\ell q} \over \partial x^k} -  {\partial
g_{k q} \over \partial x^\ell}- {\partial
g_{k \ell} \over \partial x^q} \bigr\} . \eqno(21)$$
This induced connection form, in the formula for curvature, would give
a curvature corresponding the the Hamilton manifold.  However, it is
antisymmetric in its lower indices $(\ell,k)$ (torsion).  Taken along a line
parametrized by $t$, corresponding to geodesic motion,  the
antisymmetric terms cancel, leaving precisely the symmetric connection
form $(19)$ \footnote{*}{Note that since $(19)$ and $(21)$ are not
 directly derived from $g_{ij}$, they are not metric compatible
 connections. However, performing parallel transport on the local flat tangent
 space of the Gutzwiller manifold, the resulting connection, after
 raising the tensor index to reach the Hamilton manifold, results in
 exactly the ``truncated'' connection $(19)$.} .
 A complete discussion of the tensors on the Gutzwiller
and the Hamilton manifolds  will be given elsewhere. We note here,
however, that the curvature associated
with the geodesic deviation in the Hamilton manifold, as we shall see
below, is not the same
as the intrinsic curvature of that manifold, determined by
$\Gamma_{H\ell k}^m$, 
 but rather, due to the
presence of torsion, a special
curvature form associated with the geodesics themselves.
\par Since the coefficients $M^\ell_{mn}$ constitute a connection
form, they can be used to construct a covariant derivative. It is this
 covariant derivative which must be used to
compute the rate of transport of   the geodesic deviation 
$\xi^\ell = x'^\ell - x^\ell$ along
 the (approximately common) motion of neighboring orbits in the
 Hamilton manifold, since it follows the geometrical structure of the 
geodesics.
\par The second order geodesic deviation equations
 \footnote{**}{Substituting the conformal metric $(7)$ into $(22)$,
and taking into account the constraint that both trajectories
$x'^\ell$ and $x^\ell$ have the same energy $E$, one sees that $(22)$
becomes the orbit deviation equation based on $(14)$.}
$$ {\ddot \xi}^\ell =  -2 M^\ell_{mn}{\dot x}^m {\dot \xi}^n - {\partial
M^\ell_{mn}\over \partial x^q} {\dot x}^m{\dot x}^n \xi^q ,\eqno(22)$$
obtained from $(18)$, can be factorized in terms of this covariant
derivative,
$$ \xi^\ell_{;n} = {\partial \xi^\ell \over \partial x^n} +
M^\ell_{nm}\xi^m. \eqno(23)$$
One obtains
$$ {{D_M}^2 \over {D_M} t^2} \xi^\ell  = {{R_M}^\ell}_{qmn} {\dot x}^q{\dot
x}^n \xi^m,\eqno(24)$$
where the index $M$ refers to the connection $(19)$, and
 what we shall call the {\it dynamical curvature} is
given by
$$ \eqalign{{{R_M} ^\ell}_{qmn} &= {\partial M^\ell_{qm} \over \partial x^n}
-{\partial M^\ell_{qn} \over \partial x^n}\cr
&+ M^k_{qm}M^\ell_{nk} - M^k_{qn}M^\ell_{mk}.\cr}\eqno(25)$$
 This
expression, as remarked above,  is  not the curvature of the
Hamilton manifold (given by this formula with $\Gamma_{H qm}^\ell$ in
place of $M_{qm}^\ell$), 
but a dynamical curvature which is appropriate for geodesic
motion.
\par   We give in the following a general formula for the geodesic deviation
in the Hamilton manifold in two dimensions, and then
show results of computer simulation for Poincar\'e plots
showing a correspondence with the prediction of instability from the
geodesic deviation.
\par With the conformal metric in noncovariant form $(7),(8)$, the
dynamical curvature
$(25)$ can be written in terms of derivatives of the potential $V$,
and the geodesic deviation equation $(24)$ becomes
$$ {D_M^2{\bf \xi} \over D_M t^2} = - {\cal V}P{\bf \xi},
\eqno(26)$$
where the matrix ${\cal V}$ is given by
$$ {\cal V}_{\ell i} =  \bigl\{ {3 \over M^2v^2} {\partial V
\over \partial x^\ell} {\partial V \over \partial x^i} + {1 \over M}
{\partial^2 V \over \partial x^\ell \partial x^i} \bigr\}. \eqno(27)$$
and
$$P^{ij} = \delta^{ij} - {v^i v^j
 \over v^2},\eqno(28) $$
with $ v^i \equiv {\dot x}^i$, defining a projection into a direction
orthogonal to $v^i$.
\par We then find for the component orthogonal to the motion
$$ {D_M^2 ({\bf v}_\perp \cdot {\bf \xi}) \over D_M t^2}=
-\bigl[ \lambda_1 \cos^2 \phi + \lambda_2 \sin^2 \phi\bigr]
({\bf v}_\perp \cdot {\bf \xi}^\ell) \eqno(29)$$
where $\lambda_1$ and $\lambda_2$ are eigenvalues of the matrix ${\cal
V}$,
and $\phi$ is the angle between ${\bf v}_\perp$ and the eigenvector
for $\lambda_1$.
\par Instability should occur if at least one of the eigenvalues of
 ${\cal V}$ is negative,
 in terms of the second covariant derivatives of the
transverse component of the geodesic deviation.
\par One may easily verify that the oscillator potential is predicted
 to be stable. Our criteria imply that the Duffing oscillator (without
perturbation, not a chaotic system) clearly indicates instability in
a neighborhood of the unstable fixed point. The potentials discussed
by Oloumi and Teychenne$^6$ also
demonstrate the effectiveness of our procedure; our results in these
cases are in agreement with theirs. The relation $(29)$ provides a
clear indication of the local regions of instability giving rise to
chaotic motion in 
  the H\'enon-Heiles model (this result will be discussed in detail elsewhere).
\par  We take for a simple illustration here a slight modification of
the fifth order expansion of a two body Toda lattice Hamiltonian (for which the
fourth order expansion coincides with the H\'enon-Heiles model)
$$ V(x,y) = {1 \over 2}(x^2 + y^2) + x^2y - {1 \over 3}y^3 + {3 \over
2} x^4 + {1 \over 2} y^4. \eqno(30)$$
This provides a new Hamilton chaotic
system for which our criterion gives a clear local signal for the
presence of instability. Fig.1 shows that the region of negative
eigenvalues does not penetrate the
 physically accessible region for $E= 1/6$;
 fig. 2 shows a Poincar\'e plot in the $y, p_y$ plane for this case, indicating
completely regular orbits.  In fig. 3, the distribution of
negative eigenvalues for $E=3$ is shown to penetrate deeply into the
physical region, and fig. 4 shows the corresponding Poincar\'e plot
displaying a high degree of chaotic behavior. The
criterion for instability we have given depends sensitively on the
energy of the system.  The critical energy for which the negative
eigenvalues begin to penetrate the physically accessible region, in
this example, is $E\cong 1/5$.
\par  The condition implied by the geodesic deviation
 equation $(26)$, in terms of
covariant derivatives, in which the orbits are viewed
geometrically as geodesic motion, is a  new condition for
instability, based on the underlying geometry, for a Hamiltonian
system of the form $(6)$. This geometrical picture
of Hamiltonian dynamics provides, moreover,  new insight into the structure
of the unstable and chaotic behavior of Hamiltonian dynamical systems.

\par We wish to thank S. Shnider, A. Belenkiy, P. Leifer, I.
Aharonovitch and Avi Gershon for helpful discussions.

\bigskip
\noindent {\it References}
\smallskip
\frenchspacing
\item{1.} C.G.J. Jacobi, {\it Vorlesungen \"uber Dynamik}, Verlag
G. Reiner, Berlin (1844); J.S. Hadamard, J. Math. Pures et Appl. {\bf
4}, 27 (1898). 
\item{2.} M.C. Gutzwiller, {\it Chaos in Classical and Quantum
Mechanics}, Springer-Verlag, New York (1990). See also W.D. Curtis and 
F.R. Miller, {\it Differentiable Manifolds
and Theoretical Physics}, Academic Press, New York
(1985), J. Moser and E.J. Zehnder, {\it Notes on Dynamical
Systems}, Amer. Math. Soc., Providence (2005), and  L.P. Eisenhardt, {\it A
Treatise on the Differential Geometry of Curves and Surfaces}, Ginn,
Boston (1909)[Dover, N.Y. (2004)].
\item{3.} V.I. Arnold, {\it Mathematical Methods of Classical
Mechanics}, Springer-Verlag, New York (1978).
\item{4.} P. Appell, {\it Dynamique des  Systemes Mecanique
Analytique},Gauthier-Villars, Paris (1953); H. Cartan,{\it Calcul
Differential et Formes Differentielle}, Herman, Paris (1967);
L.D. Landau, {\it Mechanics}, Mir, Moscow (1969).  The method was
utilized for the relativistic case by
D. Zerzion, L.P. Horwitz and R. Arshansky,
Jour. Math. Phys. {\bf 32}, 1788 (1991).
\item{5.} L. Casetti and M. Pettini, Phys. Rev. E {\bf 48}, 4320 (1993);
L. Caianai, L. Casetti, C. Clementi and M. Pettini,
Phys. Rev. Lett {\bf 79}, 4361 (1997);
L. Casetti, C. Clementi and M. Pettini, Phys. Rev. E {\bf
54}, 5669 (1996). See also, M. Szydlowski and J Szczesny, Phys. Rev. D
{\bf 50}, 819 (1994) and M. Szydlowski and A. Krawiec, Phys. Rev. D {\bf
53}, 6893 (1996), who have studied a somewhat generalized Gutzwiller form
(with the addition of a scalar potential) which accommodates the
Jacobi metric.
\item{6.} Atta Oloumi and Denis Teychenne, Phys. Rev.
E{\bf 60}, R6279 (1999).
\vfill
\eject
\bigskip
\centerline{FIGURE CAPTIONS}
\bigskip
\leftline{Figure 1.}
\smallskip
\par The dark area shows the region of negative eigenvalues for the
matrix ${\cal V} $. The light area corresponds
to physically allowable motion for  $E= 1/6$. The region of negative
 eigenvalues does not penetrate the physically accessible region in this case.
\bigskip
\leftline{Figure 2.}
\smallskip
\par A Poincar\'e plot in the $(y,p_y)$ plane for $E= 1/6$, indicating
 regular motion.
\bigskip
\leftline{Figure 3.}
\smallskip
\par The dark area of negative eigenvalues for the matrix ${\cal V}$ is
seen to penetrate deeply into the light region of physically
allowable motion for $E=3$.
\bigskip
\leftline{Figure 4.}
\smallskip
\par A Poincar\'e plot in the $(y,p_y)$ plane for $E=3$, indicating 
strongly chaotic behavior.
\vfill

\bye
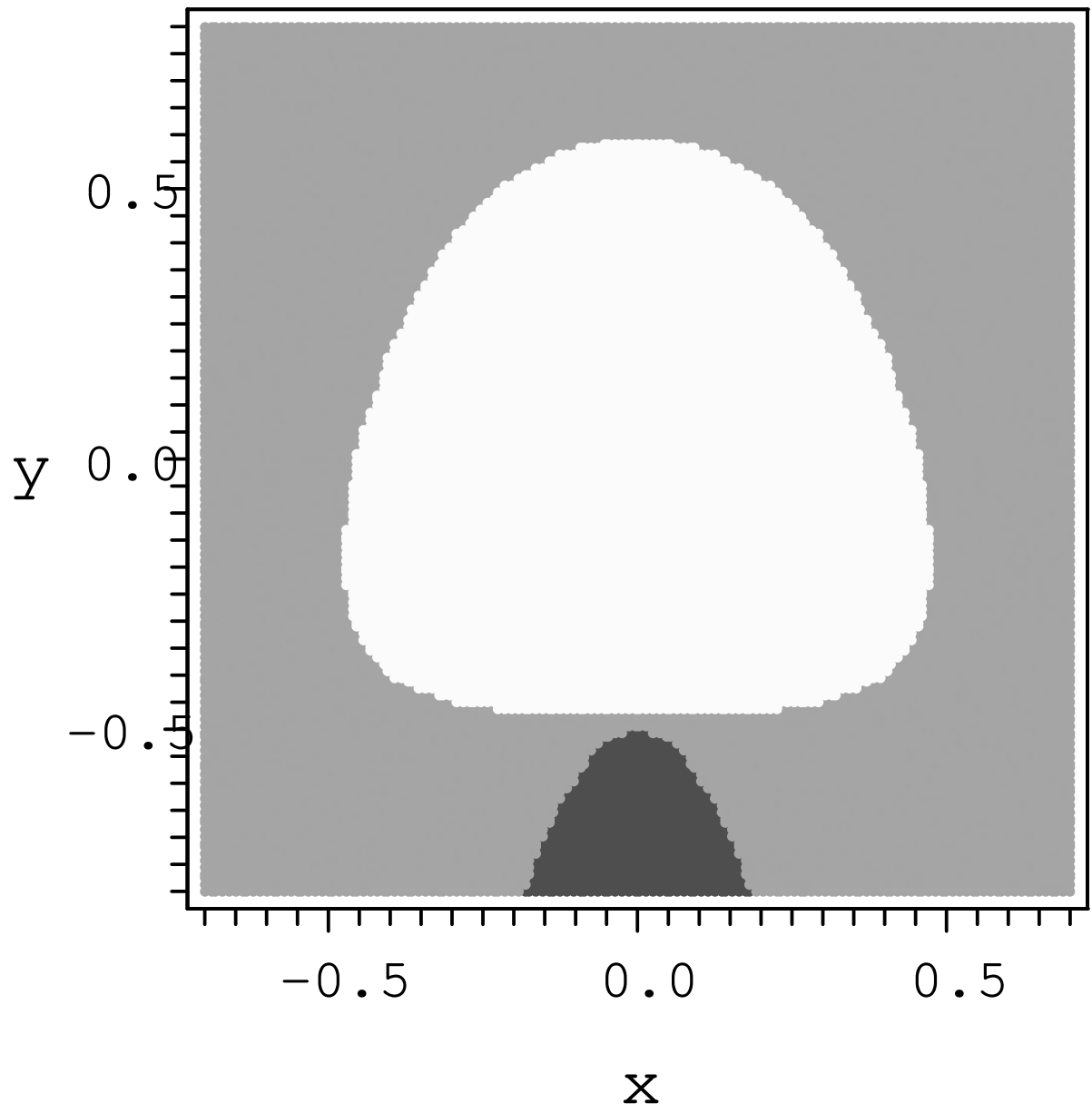
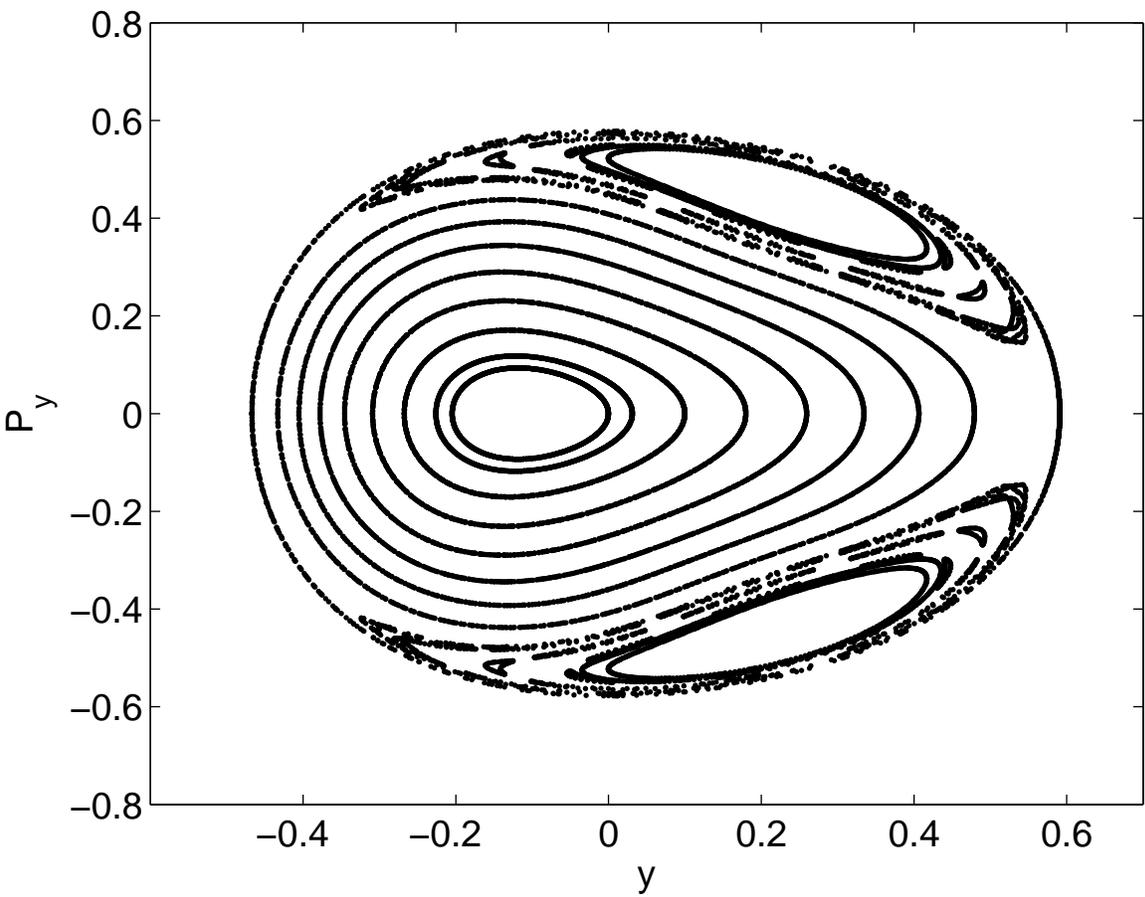
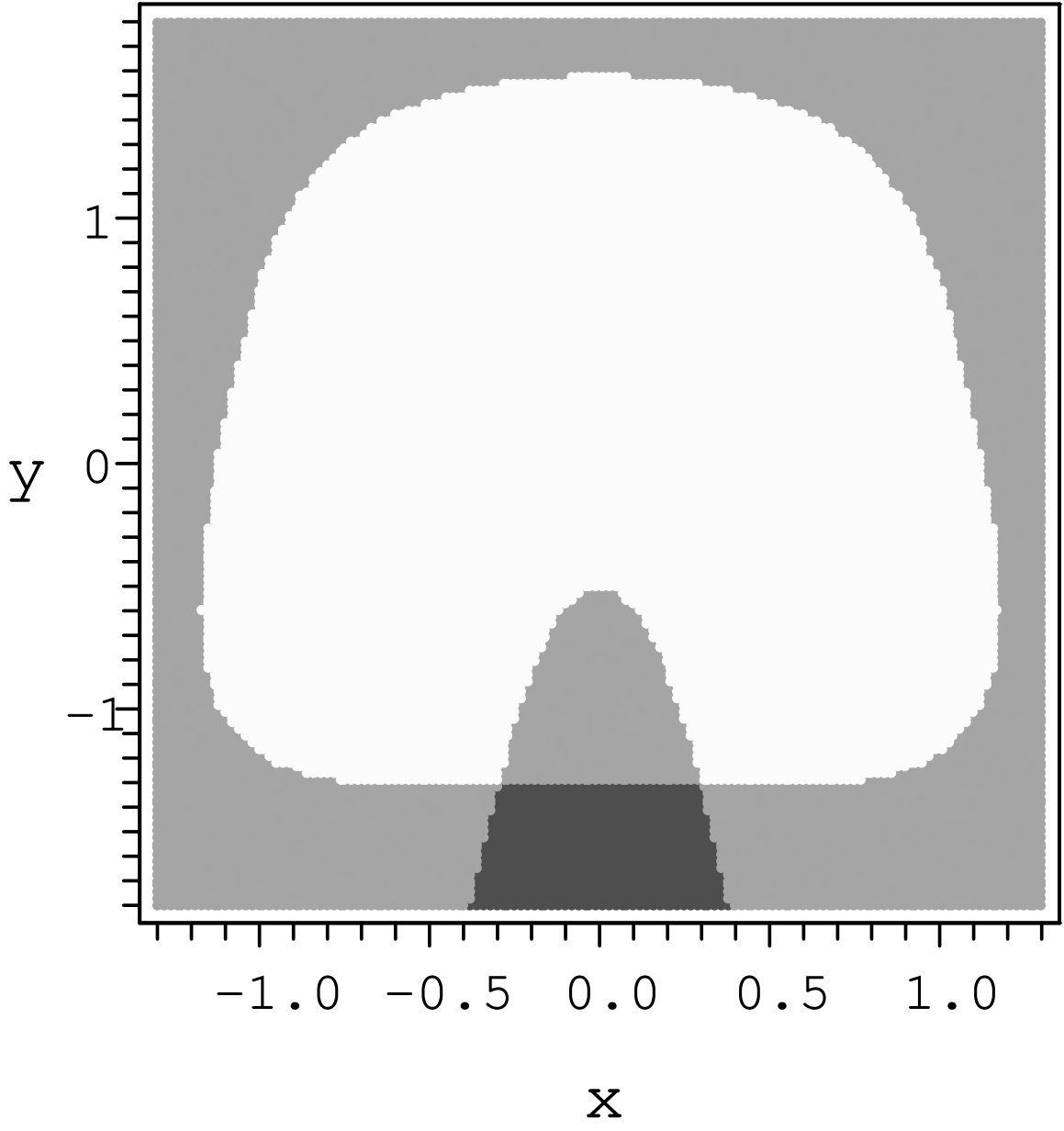
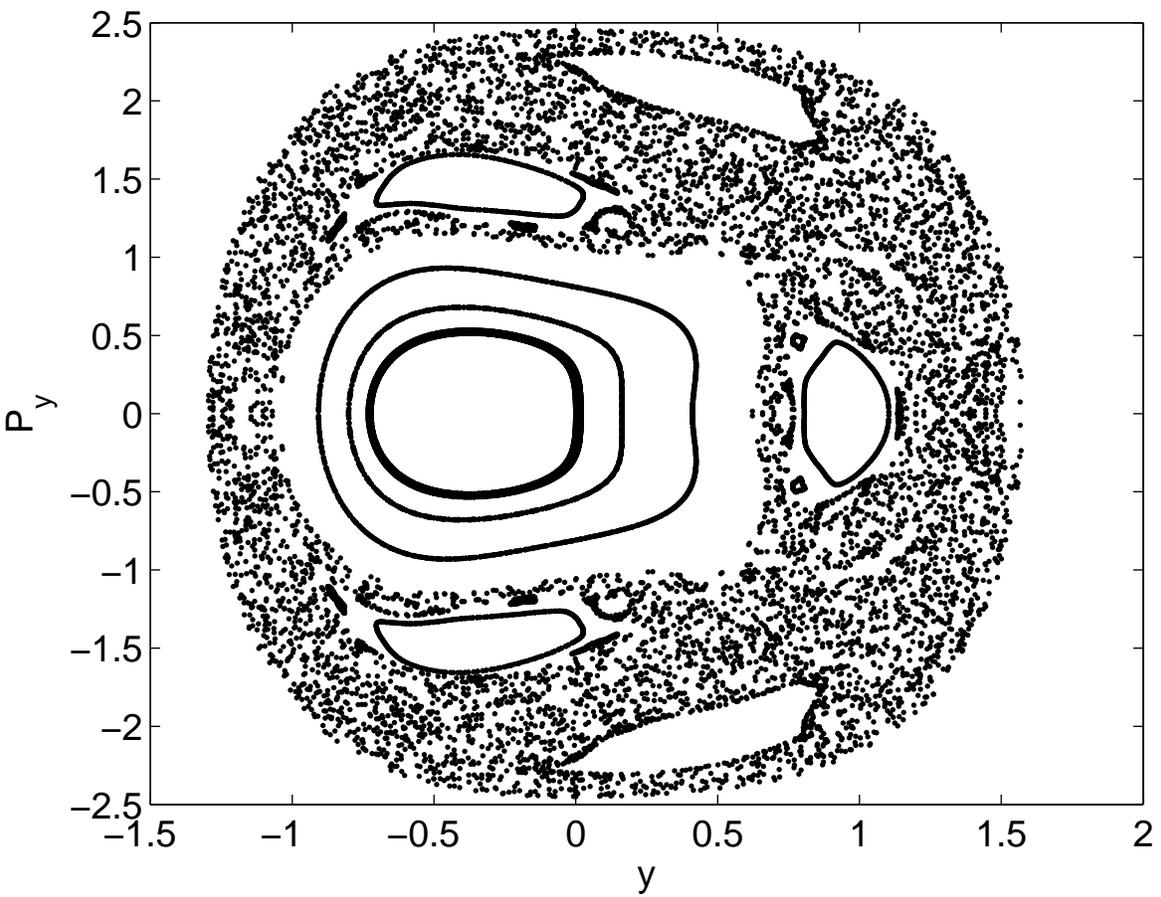
\end